\documentstyle[12pt,aaspp4,epsf]{article}
\def\fun#1#2{\lower3.6pt\vbox{\baselineskip0pt\lineskip.9pt
  \ialign{$\mathsurround=0pt#1\hfil##\hfil$\crcr#2\crcr\sim\crcr}}}
\def\lap{\mathrel{\mathpalette\fun <}}
\def\gap{\mathrel{\mathpalette\fun >}}
\def\etal{{\it et al.}}

\def\mass{{\cal M}}

\def\Msolar{{\mass_\odot}}

\lefthead{ }
\righthead{Elliptical Galaxies}

\begin{document}

\title{Dynamical Evolution of Elliptical Galaxies with Central Singularities}

\author{David Merritt and Gerald D. Quinlan}
\affil{Department of Physics and Astronomy, Rutgers University,
New Brunswick, NJ 08855}

\bigskip
\centerline{Rutgers Astrophysics Preprint Series No. 212}
\bigskip

\begin{abstract}
We study the effect of a massive central singularity on the 
structure of a triaxial galaxy using $N$-body simulations.
Starting from a single initial model, we grow black holes with 
various final masses $M_h$ and at various rates, ranging from impulsive 
to adiabatic.
In all cases, the galaxy achieves a final shape that is nearly spherical 
at the center and close to axisymmetric throughout.
However, the rate of change of the galaxy's shape depends strongly 
on the ratio $M_h/M_g$ of black hole mass to galaxy mass.
When $M_h/M_g\lap 0.3\%$, the galaxy evolves in shape 
on a timescale that exceeds $10^2$ orbital periods, or roughly a 
galaxy lifetime.
When $M_h/M_g\gap 2.5\%$, the galaxy becomes 
axisymmetric in little more than a crossing time.
We propose that the rapid evolution toward axisymmetric shapes 
that occurs when $M_h/M_g\gap 2.5\%$ provides a negative feedback mechanism 
which limits the mass of central black holes by cutting off their 
supply of fuel.

\end{abstract}

\section {Introduction}

Massive compact objects appear to be ubiquitous components of 
galactic nuclei (\cite{kor95}).
While these mass concentrations need not be black holes -- 
current observations fail by orders of magnitude to resolve the 
Schwarzschild radius -- there is a growing concensus that the 
black hole model is reasonable and, in many ways, simpler 
and more compelling than its alternatives.
Particularly attractive is the possibility that we are observing 
the black holes that once powered quasars and that still provide 
the energy source of active galactic nuclei.
This hypothesis is at least crudely consistent with the observed 
properties of quasars and AGN; in particular, quasar luminosities 
imply accumulated masses of order $10^8\Msolar$ or more per galaxy 
(\cite{sol82}), in approximate agreement with the masses inferred 
from the nuclear kinematics of nearby galaxies.
The strongest of the current black hole candidates have masses ranging 
from $\sim 10^{6.3}\Msolar$ in the case of the Milky Way to 
$\sim 10^{9.5}\Msolar$ in the case of M87; expressed as a fraction 
$M_h/M_g$ of the stellar mass of their host galaxies, black hole 
masses range from $\sim 0.02\%$ to $\sim 2\%$ (\cite{kor95}).

Central singularities containing such a large fraction of a 
galaxy's mass should have a substantial influence on the dynamics 
and evolution of the surrounding stars and gas.
Peebles (1972) and Young (1980) noted that the 
growth of a black hole in a pre-existing stellar core would 
pull the surrounding stars into a power-law density cusp.
Power-law cusps are now known to be universal components of early-type 
galaxies (\cite{geb96}), although they are often found to be
less steep than the $\rho\propto r^{-3/2}$ dependence that 
follows from adiabatic growth of a black hole growth in an 
isothermal sphere (\cite{mef95}).
Another consequence of the slow growth of a black hole is a 
polarization of the stellar velocity field:
orbits of stars near the black hole become slightly more circular
as a consequence of its growth (\cite{you80}; \cite{gob84}).
Such a polarization may have been observed in M87 (\cite{meo97}).

Both of these dynamical processes were first studied in the context of 
spherical galaxy models, and both are predicted to strongly influence the 
stellar distribution only within the radius of influence of the black hole, 
$r_h=\sqrt{GM_h/\sigma^2}$, where $\sigma$ is the velocity 
dispersion of stars in the nucleus.
However, the influence of a massive central singularity in a 
non-axisymmetric galaxy can extend far beyond $r_h$.
Many of the orbits in triaxial potentials are boxes, orbits that 
pass arbitrarily close to the center after a sufficiently long time.
A number of authors have suggested that the scattering and redistribution 
of stars on box orbits by a central black hole would cause at least the inner 
parts of a triaxial galaxy to become rounder or more axisymmetric 
(\cite{lan83}; \cite{geb85}; \cite{ger87}). 
This prediction was first tested in a self-consistent way by 
Norman, May \& van Albada (1985), who performed $N$-body integrations 
of triaxial galaxy models in which a central point mass was gradually grown.
They found a modest evolution toward sphericity at the centers of 
their models.
However, the small number of particles used by Norman, May \& van 
Albada in their simulations, coupled with the relatively large softening 
lengths that they were forced to adopt, made it difficult for them to 
interpret their results in an unambiguous way.
More recent $N$-body studies of triaxial systems by Udry (1993), 
Dubinski (1994) and others -- while based on improved algorithms
and larger numbers of particles -- have likewise lacked the temporal 
and spatial resolution needed to accurately reproduce the motion of 
stars very near to the central point mass.

The geometry of barred spiral galaxies is slightly easier to 
treat than that of elliptical galaxies, both because bars are 
essentially two-dimensional objects and because their densities 
are not very strongly peaked toward the center.
A number of $N$-body studies of barred galaxies
(\cite{frb93}; \cite{fri94}; \cite{nos96})
have shown that a bar can be essentially destroyed -- 
that is, converted into a nearly axisymmetric, bulge-like component -- 
by a central mass concentration containing a few percent 
of the total disk mass.
Test-particle integrations in barred or triaxial potentials
(\cite{pfe84}; \cite{pfz89}; \cite{mis89}; \cite{han90}; \cite{hap93}) 
indicate that the primary physical mechanism responsible for bar 
dissolution is chaos induced in the stellar motions by the central mass.
Attempts to construct self-consistent nonaxisymmetric systems 
by orbital superposition often fail if only regular, i.e. 
non-chaotic, orbits are included (\cite{sch93}; \cite{mef96}; 
\cite{kac96}; \cite{mer97b}).

In this paper, we present an $N$-body study of the 
influence of a growing black hole on a pre-existing triaxial 
galaxy.
Our study benefits to a great degree from advances in the art of 
$N$-body simulation since the time of 
Norman, May \& van Albada's (1985) pioneering work.
Our computer code, described in Section 2, employs individual 
time steps for the particles.
In this way we are able to accurately follow the stellar 
trajectories over a great range of spatial scales, to well within 
the radius of influence of the black hole.
We also assign different masses to particles with different 
binding energies in order to increase the effective resolution 
near the black hole.
Our integrations employ enough particles, $\sim 10^5$, that 
collisional relaxation effects are negligible.

Our results (Sections 3,4) are in qualitative agreement with the predictions of 
Gerhard \& Binney (1985) and others: the black hole causes the 
initially triaxial galaxy to become rounder near the center and 
more axisymmetric throughout.
But we find that the evolution toward axisymmetry is nearly 
complete; i.e. the model reaches an equilibrium state that 
differs only negligibly from (oblate) axisymmetry over a wide 
range in radii.
Furthermore we find that the timescale over which the galaxy's 
shape evolves is strongly dependent on the mass of the black hole.
When $M_h/M_g$ increases above $\sim 2\%$, the evolution to an 
axisymmetric state takes place in a time not much longer than the 
crossing time; while for $M_h/M_g\lap 0.01$, the evolution requires of order 
$10^2$ crossing times or longer.
We argue, based on test-particle integrations, that the sudden 
decrease in the galaxy response time when $M_h/M_g$ exceeds 
$\sim0.02$ is a consequence of a transition to global stochasticity in the 
phase space of the boxlike orbits, orbits that are necessary for 
maintaining triaxiality.

Our results suggest that there is a characteristic mass fraction, 
of order $M_h/M_g\approx0.02$, above which a central black hole 
will cause its host galaxy to evolve rapidly to an axisymmetric state.
This mass fraction is close to the maximum value of $M_h/M_g$ 
observed in nearby galaxies: several galaxies have 
$M_h/M_g\approx 10^{-2}$, and the current record-holders, NGC 4342 
and NGC 3115, both have $M_h/M_g\approx 0.025$.
We suggest that this agreement is more than a coincidence.
The fueling of massive black holes in AGN and 
quasars requires matter to be channeled into the nucleus from 
large distances, and gravitational torques resulting from a 
bar-shaped or triaxial mass distribution are an efficient way to 
accomplish this.
The masses of nuclear black holes may be regulated by a 
negative-feedback mechanism that shuts off the supply of fuel to 
the nucleus once $M_h/M_g$ exceeds roughly the critical value 
found here (Section 5).

\section{Numerical Techniques and Initial Conditions}

The $N$-body code used here was a hybrid, combining elements 
from the ``mean field'' code of Hernquist \& Ostriker (1992) and the {\tt 
NBODY} programs of Aarseth (1994).
The equation of motion for the stars was
\begin{equation}
{\bf \ddot r}_k = -{M(t)\ {\bf r}_k\over (r_k^2 + 
\epsilon^2)^{3/2}} - \sum_{n,l,m}A_{nlm}\nabla\Phi_{nlm}\ ;
\end{equation}
the black hole, of mass $M(t)$, remained fixed at the origin.
The first term of Equation 1 describes the interaction of the stars with the 
black hole, while the second represents the star-star 
interactions. 
The latter were computed via a basis-function expansion of the 
potential, with coefficients $A_{nlm}$ that were periodically 
updated.
The indices $n$ and $l$ run from $0$ to $n_{max}$ and $0$ to 
$l_{max}$; we used $n_{max}=16$ radial basis functions and 
$l_{max}=4$ angular basis functions.
Some test experiments with larger $n_{max}$ and $l_{max}$ produced 
very similar results.
Stars were advanced with individual timesteps (sometimes 
differing by as much as a factor of $10^6$) using Aarseth's (1994) 
fourth-order {\tt NBODY1} integrator.
The timesteps were updated after each step by a function 
involving the force and the first three time derivatives of the 
force, multiplied by the square root of an accuracy parameter 
$\eta$.
We used $\eta=0.01$, smaller than the value recommended by 
Aarseth $(\eta=0.02-0.03)$.
The coefficients $A_{nlm}$ for the potential expansion were 
updated at fixed time intervals $\Delta t = 0.0005$, after the 
coordinates of the stars were predicted to a common time.

The same initial model was used for all of the integrations 
described here.
We generated a steady-state, triaxial, nonrotating $N$-body galaxy 
containing $1.1\times 10^5$ particles as follows.

1. A random, spherical distribution of $10^4$ particles with density 
profile $\rho\propto r^{-1.5}$, $0<r<2$ was set up. 
The particles were initially at rest.
The combined mass of these particles was set equal to one; 
fixing $G=1$ thus determined the model units.

2. This configuration was evolved forward with the $N$-body code, 
with $M_h=0$, until a time $t=50$.  During this time, the model collapsed 
and re-expanded, forming in the process a nearly steady-state, triaxial 
bar as in Aguilar and Merritt (1990) and Cannizzo \& Hollister 
(1992).

3. The particles in the equilibrium triaxial model were sorted by binding 
energy into 10 groups.
Each particle in the most-bound group was replaced by 10 
particles, whose coordinates were computed by integrating the 
triaxial model for 2 additional times units and storing the 
positions and velocities at 10 equally-spaced times.
In the same way, the particles in the next binding-energy group 
were replaced by 9 particles each, then by 8, etc.; particles in 
the least-bound group were not subdivided.
The total mass of the model was left unchanged.

This scheme for subdividing particles was designed to increase the 
resolution of the $N$-body code very near the center, where the 
black hole produces a cusp in the density profile.

4. Each of the $5.5\times 10^4$ particles so generated was then 
replaced by two particles, one at the original position $({\bf 
r}, {\bf v})$ and one at $(-{\bf r}, -{\bf v})$.  
This procedure guaranteed that the center of mass of the model 
was initially at the origin and that the model had no net 
momentum.

5. The resulting set of $1.1\times 10^5$ particles was evolved 
for another 20 time units to ensure that the model had reached 
equilibrium.
The configuration obtained at the end of this integration was 
defined as the initial model.

The initial model is a triaxial body with a short-to-long 
axis ratio $c/a$ of approximately 0.5, and a ``triaxiality index" 
$T=(a^2-b^2)/(a^2-c^2)\approx 0.6$.
The model is strongly peanut-shaped as seen from the 
intermediate axis (Figure \ref{fig5}), a common feature of $N$-body bars.
The spherically-symmetrized properties of the model are shown 
in Figure \ref{fig3}.
The density profile exhibits a constant-density core of radius 
$r_c\approx 0.08$, and a half-mass radius of 0.48.
The particle motions are approximately isotropic in the core, 
and become increasingly radial in the envelope (Figure \ref{fig3}c) -- 
a relic of the radial collapse.
Table 1 lists some of the global properties of the initial model.
Roughly 89\% of the mass in the original spherical cloud remained 
bound after the collapse.
(Thus the bound mass of the initial triaxial model is 0.89 in the units 
adopted here.)
Approximating the model as spherically symmetric, the full period 
of a circular orbit at the radius containing one-half of the 
bound mass is $\sim 2.9$.

Figure \ref{fig1} shows that the initial model is close to 
equilibrium.
Plotted there is the time dependence of the axis ratios of the 
most-bound 2\%, 10\% and 50\% of the particles, defined as the 
axis ratios of a homogeneous spheroid with the same 
moment-of-inertia tensor as the particles.
Over an elapsed time of 50 units -- roughly 17 half-mass orbital periods -- 
there is no discernible change in the shape of the model.
The radii containing 1\%, 2\%, etc. of the total mass were
likewise found to remain essentially constant over this interval.

The black hole was grown by increasing the mass at the center 
of the model according to the relation
\begin{eqnarray}
M(t) & = & M_h \tau^2 (3-2\tau), \ \ \tau\le 1;\nonumber \\
& = & M_h, \ \ \tau > 1,
\end{eqnarray}
with $\tau=t/t_{grow}$ (Figure \ref{fig2}).
Various values for $M_h$ and $t_{grow}$ were used, as discussed 
below.
Because the mass of stars bound to the initial model is 
0.89 in our units, the final black hole mass 
-- expressed as a fraction of the bound mass of the galaxy -- is 
approximately $1.12 M_h$.

The black hole softening length $\epsilon$ of Equation (1) was 
computed via the relation
\begin{equation}
v_{\epsilon}=8=\sqrt{GM_h/\epsilon};
\end{equation}
in other words, the softening length was chosen such that a
typical star at a distance $\epsilon$ from the 
black hole would have an orbital velocity of 8 in model units.
For comparison, the central velocity dispersion of the initial 
model is approximately 1.2 (Figure \ref{fig3}b).
The softening length -- which is a function of $M_h$ -- is  
given in Table 1 for the three values of $M_h$ used here.
Our adopted softening length was always much smaller than the 
radius of influence $r_h$ of the black hole.
By choosing $\epsilon\ll r_h$, we were able to accurately 
reproduce the formation of a steep stellar cusp as the black hole 
grew.

The $N$-body code conserved the total energy of the model to 
$|\Delta E/E|\approx 0.002$ over 50 time units with no black hole.
Once the black hole had grown, energy was typically conserved to 
$|\Delta E/E|\approx 0.02$ over the same time interval.

\section {Results}

The growth of the black hole caused the initially triaxial
model to become nearly spherical near the center.
We therefore begin by describing the spherically-symmetrized structure 
of the models at late times, and we compare our results with those 
of earlier modelling that imposed spherical symmetry.
Here and below, smooth estimates of spatially-dependent 
properties were computed from the discrete particle coordinates
via nonparametric function estimators, 
either adaptive kernels or smoothing splines 
(\cite{wah90}; \cite{grs94}).

\subsection{Spherically Symmetrized Profiles}

Figure \ref{fig3}a shows the dependence of the central density profile 
at $t=40$ on $M_h$, for three integrations with $t_{grow}=15$.
The initial model has a small core, with radius $r_c\approx 
0.08$.
This core is replaced by a cusp following the growth of the 
black hole in all three models; within the cusp, the dependence 
of density on radius is well approximated as a power law, 
$\rho\sim r^{-\gamma}$ with $\gamma\approx 2$.
In the runs with the smaller black holes, $M_h=0.003$ and 
$0.01$, the signature of the core persists even after growth of 
the black hole in the form of an inflection in the density profile 
at $r\approx r_h$.

Quinlan, Hernquist \& Sigurdsson (1995) computed the evolution 
of the stellar distribution function $f(E,L^2)$ of a spherical model 
as the mass of a central singularity was slowly increased.
In models with cores, the growth of the black hole produced 
profiles similar to those of Figure \ref{fig3}a (e.g. their Fig. 3);
the logarithmic slope of the central density cusp was usually 
close to $-2$ and only weakly dependent on the initial 
density profile.
Their results were confirmed by Sigurdsson, Quinlan \& Hernquist 
(1995) in spherically-symmetric $N$-body integrations.

The velocity dispersion profile (computed by assuming a spherical 
velocity ellipsoid) also exhibits a cusp after the growth of 
the black hole (Figure \ref{fig3}b), with $\sigma\sim r^{-1/2}$, as expected 
for particles moving in response to an inverse-square force law.
The velocity dispersion in the initial model falls 
slightly toward the center, a common feature in models with 
steeper-than-isothermal central density profiles (e.g. 
\cite{bin80}).
This drop is retained following the growth of the smallest 
black hole, $M_h=0.003$, and even for $M_h=0.03$ the velocity 
dipsersion profile exhibits a mild inflection at $r\approx r_h$.
These results are again similar to those of Quinlan, Hernquist \& 
Sigurdsson (1995) (e.g. their Fig. 4) and Sigurdsson, Quinlan \& 
Hernquist (1995). 

Adiabatic growth of a central point mass in an initially 
constant-density core is expected to circularize the stellar 
orbits slightly (\cite{you80}; \cite{gob84}).
The effect is subtle, since orbital eccentricities are left 
essentially unchanged by slow changes in the potential 
(\cite{lyn63}); the velocity polarization is a second-order 
effect resulting from the different character of orbits in
the Keplerian and harmonic-oscillator limits.
We nevertheless see the effect here (Figure \ref{fig3}c): the velocity 
anisotropy, $\beta(r) = 1 - \sigma_t^2(r)/\sigma_r^2(r)$, falls 
slightly below zero for $r\lap r_h$ in the models with $M_h=0.01$ 
and $0.03$, indicating a slight bias toward circular motions.
However, $\beta$ does not drop below about $-0.2$.
Quinlan, Hernquist \& Sigurdsson (1995) and Sigurdsson, Quinlan 
\& Hernquist (1995) found similar, modest changes in $\beta$
at small radii in spherical models with comparable black hole masses.

The ``observable'' velocity dispersion profile is presented in 
Figure \ref{fig3}d.
Here the averaging was carried out over circular rings in the 
plane of the sky, defined as the plane perpendicular to the 
intermediate axis of the model.
The $r^{-1/2}$ velocity cusp is clearly visible in the projected 
profile as well.

To summarize: The slow growth of a central point mass induces 
changes near the center of the $N$-body model that are similar to those 
predicted by Young (1980), Goodman \& Binney (1984), 
Quinlan, Hernquist \& Sigurdsson (1995) and Sigurdsson, Quinlan 
\& Hernquist (1995) on the basis of spherical models.
The constant-density core is replaced by an $r^{-2}$ cusp; the
velocity dispersion increases as $r^{-1/2}$ near the black hole; 
and the velocity ellipsoid becomes mildly elongated in directions
tangential to the radius vector.
All of these changes are confined to a radius $r\lap r_h$.

\subsection{Model Shapes}

Not all effects of the central singularity are confined to 
small radii, however.
Figure \ref{fig4} illustrates the time-dependence of the model axis 
ratios for three integrations with $M_h=0.003,0.01, 0.03$ and 
$t_{grow}=15$. 
Axis ratios were defined in terms of the principal axes 
$a\ge b\ge c$ of the homogenous ellipsoid with the 
same moment-of-inertia tensor as the $N$-body model.
Following the usual practice, particles were first ranked according 
to binding energy and the inertia tensor was computed for various 
subsets, e.g. the most-bound 2\%, 10\% etc.
Figure \ref{fig4} shows that the model changes shape even at large radii, 
out to and exceeding the half-mass radius.
Near the center -- at radii corresponding to the inner few 
percent of the mass in stars -- all of the models become nearly 
spherical, on a timescale that is comparable to $t_{grow}$.
At larger radii, both axis ratios $b/a$ and $c/a$ initially increase, 
but the evolution appears to slow or halt once the models 
approach axisymmetry, i.e. when $b/a$ comes close to one.
This is seen most clearly in the integration with $M_h=0.03$, where the 
evolution is essentially complete once the black hole has reached its 
full mass at $t=15$.
In the runs with the smaller black hole masses, axisymmetry 
within the half-mass radius is barely reached at the final time step 
and slow evolution continues until the end of the run.
The evolution timescale is also a function 
of location in the model, i.e. longer at larger radii.

The evolution seen here toward rounder or more axisymmetric 
shapes is not surprising; similar evolution has been observed in a number 
of $N$-body studies of triaxial galaxies following an increase in 
the central density 
(\cite{nom85}; \cite{udr93}; \cite{dub94}; \cite{bah96}; 
\cite{nos96}).
The mechanism that drives the evolution is discussed in detail 
below; here we note only that the evolution appears to be dependent 
both on the presence of a central singularity and on 
departures from axisymmetry, since it slows or stops once the 
models are close to axisymmetric.

One nevertheless worries that the evolution seen in Figure \ref{fig4} 
might be due in 
part to systematic errors in the $N$-body code.
Most errors in the force computations or the time integration would 
tend to produce spurious evolution toward spherical symmetry.
We note that the run with the largest black hole exhibits 
essentially no evolution once axisymmetry is reached; the model 
accurately maintains its highly flattened, axisymmetric shape.
On the other hand, one might not expect integration errors to 
have much effect in axisymmetric models since angular momentum 
conservation would keep most particles from approaching the black 
hole.

The fact that the stellar cusps, once formed, persist without 
significant change also argues in favor of the $N$-body 
integrations being carried out correctly in the vicinity of the 
black hole.
But here, too, a cautionary note is in order.
The representation of the potential via a basis set (Eq. 1) is 
not optimal.
Since the basis functions used here correspond to mass components 
with central cusps, the $N$-body code might tend to produce cusps or 
to spuriously maintain them once formed.

Although the timescale for the evolution in Figure \ref{fig4} is strongly
dependent on $M_h$, the equilibrium figures of the three models are 
very similar: all models tend toward axisymmetry, even at radii 
exceeding the half-mass radius, and the short-to-long axis ratio 
of the final configuration varies from $\sim 1$ at the smallest radii 
to $\sim 0.6$ near the half-mass radius in each of the models.
Figure \ref{fig5} illustrates the projected appearance of the model with the 
intermediate-mass black hole, $M_h=0.01$, at the final time step 
($t=80$), as seen from a point in the equatorial plane.
The initial model is also shown for comparison, as seen from the 
intermediate axis.
The elongation of the initial model is approximately constant 
with radius, and its isophotes are strongly peanut-shaped at 
intermediate radii,
The isophotes of the model with $M_h=0.01$ become 
progressively more elongated with increasing radius, although 
they remain rounder than those of the initial model at least 
until a radius of $\sim 5$, where the dynamical time is $\sim 
80$, roughly equal to the integration time.

Boxy or peanut-shaped isophotes like those in Figure \ref{fig5}
are a common feature of $N$-body galaxies 
(e.g. \cite{may85}; \cite{dub92}; \cite{her92}; \cite{all92}; \cite{udr93}).
By contrast, real elliptical galaxies have isophotal shapes that are almost 
precisely elliptical: the deviations are typically around 
$0.5\%$, rarely more than about $2\%$, and are as often in the 
direction of diskiness as of boxiness (\cite{ben88}).
As far as we know, no convincing explanation for the lack of 
significant boxiness in real ellipticals has yet been given.
Based on Figure \ref{fig5}, we propose that the precisely ellipsoidal 
shapes of elliptical galaxies may be a consequence of evolution 
induced by nuclear black holes.

\subsection{Evolution Timescales}

Inspection of Figure \ref{fig4} reveals one intriguing result.
In the integration with the largest black hole, $M_h=0.03$, the 
model reaches axisymmetry at all three radii displayed $(r_{2}, r_{10}, 
r_{50})$ by roughly the time the black hole has ceased growing, i.e. by 
$t\approx 15$.
In other words, the evolution of the model's figure appears to be 
limited only by the growth rate of the black hole.
By comparison, in the integrations with smaller black holes, 
$M_h=0.01$ and $M_h=0.003$, the model figure evolves at different 
rates at different radii, and the evolution requires much 
longer to reach completion -- roughly 100 time units ($\sim 30$ 
orbital periods) at the half-mass radius in the model with $M_h=0.01$, 
and perhaps 200-300 time units ($\sim 80$ orbital periods) in the model 
with $M_h=0.003$.
Only at the very centers of these two models -- where the orbital 
period is a small fraction of the black hole growth time -- does 
the shape of the model appear to evolve at a rate that is limited 
by the growth of the black hole.
Outside of the very center, the model figure evolves on a 
timescale that is approximately a constant multiple (different for 
the two values of $M_h$) of the local orbital period. 

To better understand the very rapid evolution observed in the model 
with $M_h=0.03$, we repeated this integration with three different 
values of the black hole growth time: $t_{grow}=0.1, 5$ and $45$.
The shortest of these growth times, $t_{grow}=0.1$, is small 
compared even with the orbital period at the center of the 
initial model; hence the growth of the black hole in this integration 
is essentially impulsive as seen by the stars.
Figure \ref{fig6} reveals that the model responds to the 
$M_h=0.03$ black hole about as quickly as it can; in the impulsive 
limit, the model evolves to axisymmetry at all radii on a timescale
that is close to the local orbital period.
For instance, the intermediate-to-long axis ratio $b/a$ 
determined by the 50\% most-bound stars reaches unity by a time 
of $\sim 2.5$ when $t_{grow}=0.1$, compared to a circular orbital 
period of $\sim 2.9$ at the half-mass radius (Table 1).

These experiments suggest that the timescale on which the galaxy 
responds in shape to the appearance of the black hole is a steep 
function of the black hole's mass.
Figure \ref{fig7} plots the evolution toward axisymmetry of the most-bound 
50\% of the stars for the three values of $M_h$.
The curves for the two smaller black hole masses were taken from 
the simulations with $t_{grow}=15$, which is shorter than the
apparent response time of the galaxy at the half-mass radius for 
these values of $M_h$.
The curve for $M_h=0.03$ is from the integration with 
$t_{grow}=5$, short enough that the galaxy's response at the 
half-mass radius is not strongly limited by the black hole growth rate.
Figure \ref{fig7} suggests that the response time of the galaxy to the 
black hole drops suddenly when $M_h$ exceeds $\sim 10^{-2}$.

The critical mass fraction can be estimated more accurately by 
looking at the integration with the longer black-hole growth 
time, $t_{grow}=45$ (Figure \ref{fig6}a).
Axisymmetry at all radii is reached in these integrations 
somewhat before the black hole has attained its final mass of $0.03$.
For instance, at the radius containing $10\%$ of the stars, 
$b/a\approx 1$ when $t\approx 30$, at which time the black hole 
mass is $\sim 0.025$, or $M_h/M_g\approx 0.027$.
At the radius containing $50\%$ of the stars, approximate 
axisymmetry is reached when $t\approx 40$; taking into account 
the orbital time at this radius suggests a critical mass 
fraction for transition to axisymmetry of $M_h/M_g\approx 0.028$.
Thus, evolution toward axisymmetry takes place very 
rapidly once $M_h/M_g$ exceeds $\sim 2.5\%$.

\section{Evolution Mechanism}

The evolution toward axisymmetric shapes seen here is an expected 
consequence of stochasticity in the stellar orbits, induced by the 
central singularity.
The argument -- which is not original with us -- goes 
roughly as follows; fuller discussion may be found in the papers 
of Schwarzschild (1981), Norman, May \& van Albada (1985), 
Gerhard \& Binney (1986),
Gerhard (1987), Udry \& Pfenniger (1988), 
Merritt \& Fridman (1996), Barnes \& Hernquist (1996), 
Merritt \& Valluri (1996), and Merritt (1997a,b).

1. Many of the stars in the initial, triaxial model are on 
regular box orbits that carry them close to the center, once per 
orbital period. 
(A regular orbit is defined as an orbit with three isolating 
integrals of the motion, two in addition to the energy.)

2. Most of the box orbits lose two of their three integrals of 
motion once the black hole appears, essentially because of 
large-angle deflections suffered during close passages 
to the center.
(The exceptions are boxlike orbits that lie close to a stable 
resonant orbit that avoids the center, e.g. the $2:1$ ``banana''
orbit.)

3. Trajectories defined by only one integral of motion -- the 
energy --  densely fill the region within the equipotential 
surface $\Phi({\bf x})=E$.
The time-averaged shape of such an orbit is approximately 
spherical.

4. Because box orbits with a variety of shapes are crucial for 
maintaining strongly triaxial figures, the model can not preserve its 
triaxial shape once the black hole appears.
It begins to evolve toward more nearly spherical, or 
at least axisymmetric, shapes.

5. As axisymmetry is approached, the boxlike orbits disappear, 
since axisymmetric potentials can only support tube orbits -- orbits 
that conserve angular momentum about the symmetry axis
and avoid the center.
The black hole is then no longer effective at inducing 
stochasticity in the orbital motion, and the galaxy evolves to a 
steady, axisymmetric state in a few crossing times.

The evolution of the $N$-body models described above is consistent 
on a qualitative level with this picture.
Especially compelling to us is the fact that the $N$-body models 
evolve in shape only when two conditions are satisfied: the model 
is triaxial, i.e. departs from axial symmetry; and $M_h\ne 0$.
The initial, strongly triaxial model does not evolve (Figure \ref{fig1}), 
presumably because the box orbits which it contains are regular, 
or nearly so -- consistent with the expected behavior of boxlike 
orbits in triaxial models with smooth cores (e.g. \cite{gos81}).
And after the black hole is grown, the models cease evolving once 
axisymmetry is reached (Figures \ref{fig4}, \ref{fig6}), i.e. once the 
box orbits have disappeared.
Furthermore, the evolution of the model's shape is rapid compared 
to the expected rate of two-body relaxation, nor is any evidence 
of collisional relaxation apparent in the test integrations of 
the initial model without a black hole.
Taken together, these facts suggest that the evolution which we 
observe is being driven primarily by the interaction 
of boxlike orbits with the central singularity.

Further evidence in favor of this picture is provided by the 
timescales for evolution observed in the $N$-body models.
Once a black hole is introduced, the box orbits would be expected 
to respond by gradually filling the phase-space volume made 
available to them -- i.e. the volume defined by the full energy 
hypersurface, minus those regions occupied by regular orbits, 
mostly tubes.
This relaxation toward a uniform population of the accessible 
phase space is called chaotic mixing (\cite{kam94}), 
and it has a characteristic timescale that can 
be measured via test-particle integrations in fixed potentials.
Merritt \& Valluri (1996) carried out such calculations in two triaxial 
models, one with a central density cusp, the other with 
a smooth core and a central point mass containing $0.3\%$ of the 
mass of the model.
Those authors evolved ensembles of $10^4$ particles started from 
various regions on the equipotential surface and recorded their 
relaxation to a steady state.
Mixing timescales in both models were found to depend on the 
starting point of the ensemble.
Ensembles located initially in the most stochastic parts of phase 
space -- e.g. near the short axis of the figure -- were found to mix 
most rapidly, filling much of their allowed region after just $10-30$ 
orbital periods.
Other ensembles were found to mix more slowly, remaining trapped 
in limited regions of phase space for $10^2$ orbital periods or 
longer.
Merritt \& Valluri (1996) suggested an average relaxation time associated 
with chaotic mixing in these models of $\sim 100$ orbital 
periods.
This estimate is quite consistent with the rate at which the 
$N$-body model with $M_h=0.003$ is found here to evolve -- the 
intermediate-to-long axis ratio changes from $0.76$ to $0.9$ in 
about 50 half-mass orbital periods (Figure \ref{fig7}).

Merritt \& Valluri (1996) did not extend their mixing 
calculations to triaxial models with larger black hole masses, nor 
are we aware of any other 3-D test-particle studies that could be 
usefully compared to our $N$-body models with $M_h=0.01$ and 
$0.03$.
For this reason, we carried out new test-particle 
integrations in triaxial models with central point masses of 
various sizes.
We adopted the mass model of Merritt \& Valluri (1996), which has 
density
\begin{equation}
\rho(m) = {\rho_0m_0^2\over (1+m^2)(m_0^2+m^2)}, \ \ 0\le m_0\le 
1;
\end{equation}
the central density $\rho_0$ is related to the total mass $M$ by 
$\rho_0=M(1+m_0)/(2\pi^2abcm_0^2)$.
The parameter $m_0$ is a core radius; we chose $m_0=0.1a$.
The axis ratios were taken to be $c/a=0.5$ and $b/a=0.79$, similar
to the values in the initial $N$-body model.
To this model was added a central point mass containing various 
fractions $M_h$ of the total mass $M$; we used 
$M_h=0,0.003,0.01,0.03$, and $0.1$.
Test-particle integrations were then carried out for sets of 
orbits with starting points distributed over the half-mass 
equipotential surface, as described in Merritt \& Valluri (1996).

Of greatest interest here is the configuration space volume 
filled by boxlike orbits over a relatively short interval
of time.
We accordingly integrated orbits for only $10$ periods of the 
long-axis orbit.
We found a striking result: over this short time interval, the 
behavior of the boxlike orbits changes suddenly as $M_h$ is 
increased (Figure \ref{fig8}).
For $M_h\lap 0.01$, the boxlike orbits -- particularly the thin 
boxes with starting points near the long axis of the figure, 
which are crucial for maintaining a triaxial shape -- showed 
relatively little evidence of stochastic evolution. 
While often not regular, the orbits in these models filled 
regions quite similar in shape to the regions occupied by orbits 
in the model with $M_h=0$.
By contrast, when $M_h$ was increased from $0.01$ to $0.03$, essentially 
none of the boxlike orbits were found to maintain well-defined 
boxy shapes over 10 orbital periods.
Instead, they quasi-randomly filled a rougly axisymmetric region, 
with almost no hint of ``memory'' from one oscillation to the 
next of the location of the previous turning point.
The only exceptions were orbits with starting points near to the 
$x-z$ banana.

We speculate that the change in the character of the boxlike 
orbits as $M_h$ is increased from $0.01$ to $0.03$ is indicative 
of a transition to global stochasticity (e.g. \cite{lil83}).
In many dynamical systems, one observes a sharp transition as a 
parameter is varied: from stochastic motion that is closely bounded by 
KAM surfaces, i.e. regular orbits; to motion that is strongly 
interconnected over large portions of the space.
In the latter regime, stochastic trajectories at a given energy 
are nearly indistinguishable, moving rapidly -- in just a few 
oscillations -- from one part of stochastic phase space to another.
This is just the behavior illustrated in Figure \ref{fig8} as $M_h$ is 
increased above $0.01$.

Whether or not this interpretation is correct, Figure \ref{fig8} suggests 
that the timescale over which stochasticity induces changes in 
the phase-space population of a triaxial model should drop from $\gg 
10$ orbital periods when $M_h\lap 0.01$, to $\lap 10$ orbital 
periods when $M_h\gap 0.03$.
This is what we observe in the $N$-body models (Figure \ref{fig7}).
We might predict an even steeper dependence of evolution 
rates on $M_h$ in the $N$-body models than in the test-particle 
integrations, since the $N$-body models develop a 
central density cusp with a strength that
increases with $M_h$ (Figure \ref{fig3}a), and a cusp will itself induce 
stochasticity in the motion of boxlike orbits (\cite{mef96}).

One final question concerns the end states of the $N$-body models.
Their figures are remarkably close to axisymmetric.
This is particularly true with regard to the integrations with 
the largest black hole, $M_h=0.03$, where $b/a$ at the final 
time step is consistent with unity at all radii 
displayed in Figures \ref{fig4} and \ref{fig6}.
The approach to axisymmetry is not as complete in the integrations 
with the smaller black holes, but Figure \ref{fig4} suggests that a modest 
increase in the integration times would probably have produced a precisely 
axisymmetric end state for these values of $M_h$ as well.

One might have expected the $N$-body models to reach equilibrium 
without evolving so completely to axisymmetry.
For instance, Merritt (1997b), in a study of the triaxial 
self-consistency problem, found that models with $\rho\propto 
r^{-2}$ central density cusps could exist in equilibrium for 
values of $b/a$ exceeding $\sim 0.85-0.9$.
But there are a number of reasons why the $N$-body models should  
prefer even more axisymmetric states than the self-consistency
studies would suggest.
First, the $N$-body models contain both steep stellar cusps and 
central singularities; hence one would expect stochasticity to be 
more destructive of the boxlike orbits in this study than in 
that of Merritt (1997b).
Second, the self-consistency studies are relatively crude in 
their treatment of the effects of the stochasticity; for 
instance, Merritt (1997b) simply eliminated orbits that gave 
evidence of instability after 50 orbital periods.
In fact, orbital instability rates scale roughly with the 
local orbital frequency, making them a strong function of radius.
Hence one should eliminate a larger fraction of the boxlike 
orbits near the center than at large radii, making it more 
difficult to reconstruct a self-consistent triaxial shape.
Third, the rapid evolution toward spherical symmetry at the very 
centers of the $N$-body models must influence the behavior of 
orbits at larger energies, perhaps accelerating their evolution.
It is conceivable that modestly triaxial equilibria do exist even 
with central singularities, but that our scheme of growing the 
black hole in a pre-existing triaxial model prevents us from 
arriving at these end states. 

\section {Discussion}

We have found that a central singularity containing more than 
$\sim2.5\%$ the mass of its host galaxy induces a rapid 
transition from triaxiality to axisymmetry in the distribution 
of the surrounding stars, both near the center and at appreciable 
distances.
Less massive singularities induce a more gradual evolution, 
but the end state is still close to axisymmetric; in
fact, our results suggest that the shape of an initially triaxial 
galaxy containing a central singularity of mass $M_h$ is 
approximately independent of $M_h$.  
(Other features of the final state, like the central density 
profile, clearly do depend on $M_h$.  
The final state also presumably depends on the details of the initial 
model, which we did not vary).
We argued that the sudden change in the rate of evolution of the 
galaxy's shape as $M_h/M_g$ exceeds $\sim 0.025$ is due to a transition 
to global stochasticity in the phase space of the boxlike orbits,
orbits which are required to maintain triaxiality.

While it would be dangerous to generalize from such a small set of 
experiments, it seems reasonable to suppose that a sudden 
transition to axisymmetry would occur in most triaxial models as the 
mass of a central singularity is increased beyond some critical 
value.
We discuss this hypothesis in more detail below; for the moment, 
we assume that such a transition generally occurs for 
central point masses of order $0.02 M_g$.
Is there any evidence from real galaxies for such a critical mass?

It is well known that nuclear black holes often contain of order 
$1\%$ the mass of their host galaxies.
Figure \ref{fig9} shows $M_h/M_g$ for the seven galaxies with the 
largest, well-determined black hole mass fractions.
$M_g$ is defined in Figure \ref{fig9} in the usual (and, for this 
study, appropriate) way as the mass of the 
stellar bulge in the case of disk galaxies, and as the total stellar 
mass in the case of elliptical galaxies.
Uncertainties in estimates of $M_h$ vary considerably from 
galaxy to galaxy; each of the black hole masses in Figure \ref{fig9} 
is considered to be fairly accurate, with probable uncertainties of order 
$50\%$ or less.
(The galaxy NGC 4486b may have $M_h/M_g$ as great as $0.08$ but 
has been omitted from Figure 9 since the uncertainty in the mass
of its black hole is very large; in fact $M_h$ for this galaxy is 
consistent with zero (\cite{kor97})).
Figure \ref{fig9} confirms a tendency for the most massive black holes 
to contain approximately $10^{-2}$ times the stellar mass of their host 
galaxies, both in very bright (M87) and very faint (M32) galaxies.
The largest, well-determined mass fractions are associated with 
NGC 4342 and NGC 3115, which have $M_h/M_g\approx 0.026$ and 
$0.024$, respectively.
The remainder of the galaxies have $M_h/M_g\lap 0.01$.

Based on this modest sample, it appears that nature avoids making 
nuclear black holes with masses exceeding $\sim 2.5\%$ the mass 
of their host galaxy -- remarkably similar to the critical mass 
fraction that induces a transition to axisymmetry in our simulations.
We suggest that this agreement is not a coincidence.
The fueling of massive black holes in quasars and AGN requires 
matter to be funneled into the nucleus from large distances, of 
order a kiloparsec or more (\cite{gun79}).
Models that produce fuel locally, e.g. from a dense star cluster 
surrounding the black hole (\cite{vos88}), 
generally fail to supply mass at a rate adequate to power 
the most luminous active nuclei.
In order for infall to occur over kiloparsec scales, the angular
momentum of the infalling stars or gas must somehow be removed.
While many of the details of this process remain obscure, a 
number of lines of evidence suggest that large-scale asymmetries in the 
stellar distribution are a necessary ingredient (\cite{shb90}).
Gravitational torques from non-axisymmetric perturbations --
triggered by galaxy interactions, internal self-gravitational
instabilities, or both -- can remove much of the angular momentum
from the interstellar medium, allowing gas to fall into the
nucleus (\cite{nos83}).
Support for this picture comes from computer simulations of 
galaxy interactions (\cite{nog88}; \cite{her89}; \cite{bah91}); 
from the observed association of nuclear activity with asymmetric 
geometries and bars (\cite{sis80}; \cite{dah85}; \cite{mac90});
and from the fact that the host galaxies of nearby quasars
and AGN often have close, interacting companions (\cite{hun91}; 
\cite{bls91}; \cite{bae96}).

Any mechanism that enforced large-scale axisymmetry in the
stellar distribution of a galaxy would therefore be expected to 
strongly inhibit the fueling of a nuclear black hole and hence to 
limit its mass.
To be effective, such a mechanism would need to operate in
little more than a galaxy crossing time.
This is because the black holes that powered quasars must have acquired 
most of their mass in $10^9$ years or less (\cite{tur91}), and 
because the timescale for subsequent, tidally-triggered fueling is 
believed to be comparably short (\cite{bar95}).
The rapid evolution toward axisymmetry seen here when $M_h/M_g$ 
exceeds a critical value could be such a mechanism.
In this picture, nuclear black holes might be expected to often 
contain of order $10^{-2}$ times the mass of their host galaxies, and their 
masses should never appreciably exceed $0.025 M_g$ unless the galaxy 
lost a significant fraction of its stellar mass after the black 
hole formed.
An additional prediction, of course, is that galaxies containing 
``maximal'' black holes should be close to axisymmetric.

Some of the galaxies in Figure \ref{fig9} (NGC 4342, NGC 3115, NGC 4258) 
are strongly rotating, and even the slowly-rotating ellipticals 
like M87 may have formed through mergers of rapidly-rotating disks.
If so, their nuclear black holes may have grown in the pre-existing 
disks -- a rather different environment from the one modelled here.
Test-particle integrations in two-dimensional barred galaxy models have 
shown that central mass concentrations can destroy the major families of
bar-supporting orbits when $M_h/M_g$ exceeds several percent -- 
with $M_g$ here defined as the mass of the bulge plus disk
(\cite{han90}; \cite{hap93}).
This prediction has been verified in self-consistent $N$-body 
simulations of barred galaxies; for instance, Norman, Sellwood \& Hasan 
(1996) find that rapid evolution toward axisymmetry takes place when $M_h$ 
exceeds about $0.05M_g$ in a two-dimensional barred galaxy.
However, much of the orbital instability induced by central mass 
concentrations is associated with motion out of the principal planes 
(\cite{pfe84}; \cite{pff91}; \cite{mef96}; \cite{frm97}), and there are 
indications (\cite{frb93}; \cite{sem98}) that bars in fully 3D disk galaxy 
simulations are more fragile than in two dimensions, evolving rapidly in 
shape when the black hole mass fraction exceeds only $1-2\%$.
Although a great deal of work remains to be done, it is
plausible that the critical black hole mass fraction for transition to 
axisymmetry is close to $0.02$ in models with a fairly wide 
range of geometries and kinematics.

Most of the black holes in Figure \ref{fig9} fall short of the maximum mass 
proposed here; in fact, the mean value of $M_h/M_g$ among the 
roughly ten, secure black hole candidates is only $\sim0.5\%$ 
(\cite{kor95}).
While this fact does not invalidate our identification of 
$0.025M_g$ as an upper limit, we would nevertheless like to 
understand why real black holes tend to be smaller.
Following are a number of possible explanations.

1. Some host galaxies may have reached an axisymmetric state through 
processes not related to the presence of a nuclear black hole.
For instance, Raha et al. (1991) showed how coherent bending instabilities
can convert a thin bar into an axisymmetric bulge in a few crossing 
times, and Merritt (1997b) argued that a steep central density 
cusp can induce axisymmetry in much the same way that a central
singularity can.

2. A gradual transition to axisymmetry takes place in our
simulations even for black hole masses below the critical value 
(Figure \ref{fig7}).
A sufficiently slowly-growing black hole would therefore be 
expected to shut off its supply of fuel before accreting the full 
$2.5\%$ of its galaxy's mass.
``Slowly growing'' here means that the black hole gains mass on a 
timescale long compared to the orbital period at the radius
which is supplying most of the fuel.
However, it might be difficult to reconcile such small growth rates 
with the short observed lifetimes of quasars.

3. Real galaxies are not structurally identical to our $N$-body 
models; in particular, they sometimes have stellar envelopes in which
the density falls off more slowly than the $\rho\sim r^{-3.5}$ 
dependence of our models.
Adding stars at large radii would increase $M_g$ without, 
presumably, strongly influencing the response of the inner regions 
to the black hole.

4. Galaxy mergers would not change the average ratio of black hole
mass to galaxy mass, since black holes from merging galaxies
would themselves eventually merge and come to rest in the nucleus
(\cite{bbr80}; \cite{qui96}).
But galaxy mergers could greatly affect the average ratio of black hole 
mass to {\it bulge} mass, since mergers convert gas to stars and disks to 
spheroids.  
Of course, galaxy-galaxy interactions might also cause black holes to 
{\it grow} at the expense of their host galaxies by triggering infall of gas 
into the nucleus (\cite{bah91}), and there is some evidence linking 
quasar activity with galaxy interactions (\cite{bae96}).
But it seems unlikely that the masses of nuclear black holes 
have increased significantly since the quasar epoch at $z\approx 
2-3$ (\cite{ree90}).
Hence, the net effect of mergers has probably been to reduce the 
average ratio of black hole mass to bulge mass, possibly by a large factor.
One expected consequence is that the masses of black holes in elliptical 
galaxies at the current epoch should often fall below the $2.5\%$ upper limit.
This prediction is consistent with the (meager) data of Figure \ref{fig9}: 
the black holes in elliptical galaxies (M87, M32, NGC 3377) 
all have $M_h/M_g\lap 0.01$, and the largest fractional masses are seen 
in disk galaxies (NGC 4342, 3115, 4258).
However, at least one disk galaxy -- the Milky Way -- is believed to 
have a black hole mass that falls significantly below the proposed 
upper limit (\cite{ecg97}).

5. While triaxial or barlike distortions may be necessary for the 
efficient growth of black holes, they are probably not sufficient.
Transport of stars or gas into the nucleus requires a 
reduction in orbital angular momentum of roughly nine orders 
of magnitude; gravitational torques from large-scale bars
can only account for one or two of these (\cite{lpr88}).
The gas-dynamical processes that are responsible for removing the 
remaining angular momentum are poorly understood (\cite{ost93}).
The fact that most barred spiral galaxies do not exhibit significant 
nuclear activity suggests that even stellar systems with strongly 
non-axisymmetric distortions may sometimes fail to channel mass into the 
nucleus.

If a black hole's mass is close to maximal, its host galaxy 
should be nearly axisymmetric.
One can hope to falsify this prediction by looking for evidence of minor-axis 
rotation or isophote twists in the galaxies with the largest black holes.
None of the galaxies in Figure \ref{fig9} are known to show such signatures.
The most extensively modelled galaxy in Figure \ref{fig9} is M32, 
whose kinematics can be very well reproduced by axisymmetric models 
(\cite{vaz97}).

In triaxial galaxies where the black hole mass is substantially below the 
proposed upper limit, slow evolution of the galaxy's shape should 
still take place.
Based on Figure \ref{fig7}, a spheroid containing a black hole 
with $M_h/M_g\approx 1\%$ should evolve to axisymmetry in $\sim40$ 
half-mass orbital periods, roughly a galaxy lifetime for a 
typical bright elliptical.
For $M_h/M_g\approx 0.3\%$, the evolution time increases to 
$\sim 200$ orbital periods, which is still shorter than 
the lifetimes of many faint elliptical galaxies.
For instance, in M32, the circular orbital period at the half-light
radius is only about $10^7$ years (\cite{deh95}), or $\sim 0.1\%$ 
of the age of the universe.
Faint ellipticals with moderate-to-large black holes, $M_h/M_g\gap 0.3\%$,
should therefore have evolved to axisymmetric shapes by now.
(Faint ellipticals also tend to have steep central density cusps 
(\cite{geb96}), a factor which would further enhance the importance of 
orbital stochasticity in these galaxies (\cite{mer97b}).)
The Hubble type distribution of faint ellipticals, $M_B\lap -20$, 
is in fact consistent with axisymmetry, although triaxial shapes 
can not be ruled out (\cite{trm96}; \cite{ryd96}).
Bright ellipticals appear to be at least mildly triaxial as a 
class (\cite{ryd93}; \cite{trm96}) -- consistent with their 
longer dynamical times if we assume that $M_h/M_g$ for these 
galaxies is typically no greater than for the ellipticals in 
Figure \ref{fig9}.

Nuclear black holes may have acquired most of their mass in
environments very different from their current ones, before the epoch
of galaxy formation (\cite{ree92}).
If massive black holes predate galaxies, the mechanism discussed here 
would not have been effective at limiting their masses.
The discovery of nuclear black holes containing much more than $2\%$ the mass
of their host galaxies would lend support to the view that at 
least some black holes acquired most of their mass before galaxies formed.

\bigskip

This work was supported by NSF grants AST 93-18617 and AST 
96-17088 and by NASA grant NAG 5-2803.
We thank J. Sellwood, M. Valluri, H. Zapolsky and the anonymous 
referee for comments that improved the presentation.

\clearpage

\figcaption[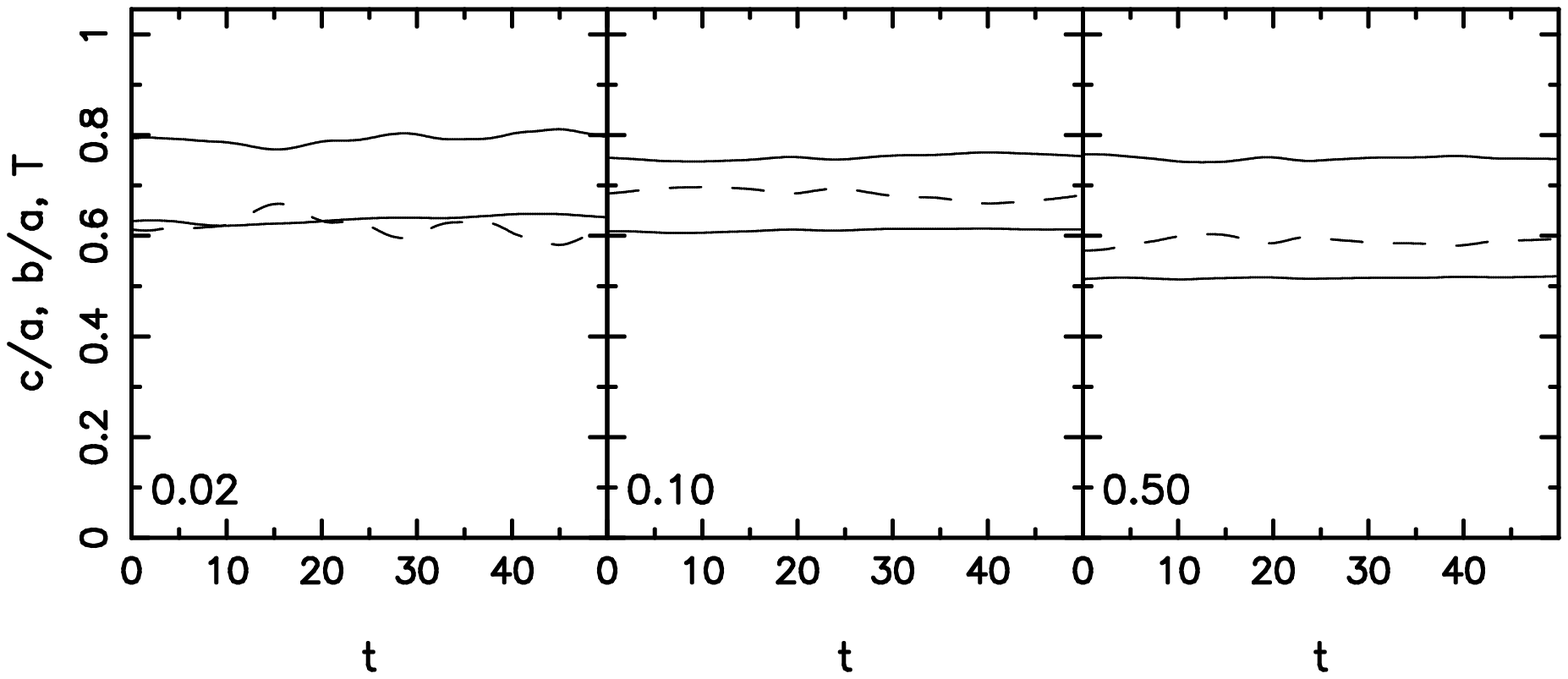]{\label{fig1}} 
Evolution of the axis ratios of the initial model in the 
absence of a black hole.
Solid lines: $b/a$ and $c/a$; dashed lines: triaxiality index, 
$T=(a^2-b^2)/(a^2-c^2)$.
The numbers in the lower left corner of each frame are the 
fraction of particles, ranked by binding energy, that were used 
in computing the axis ratios.

\figcaption[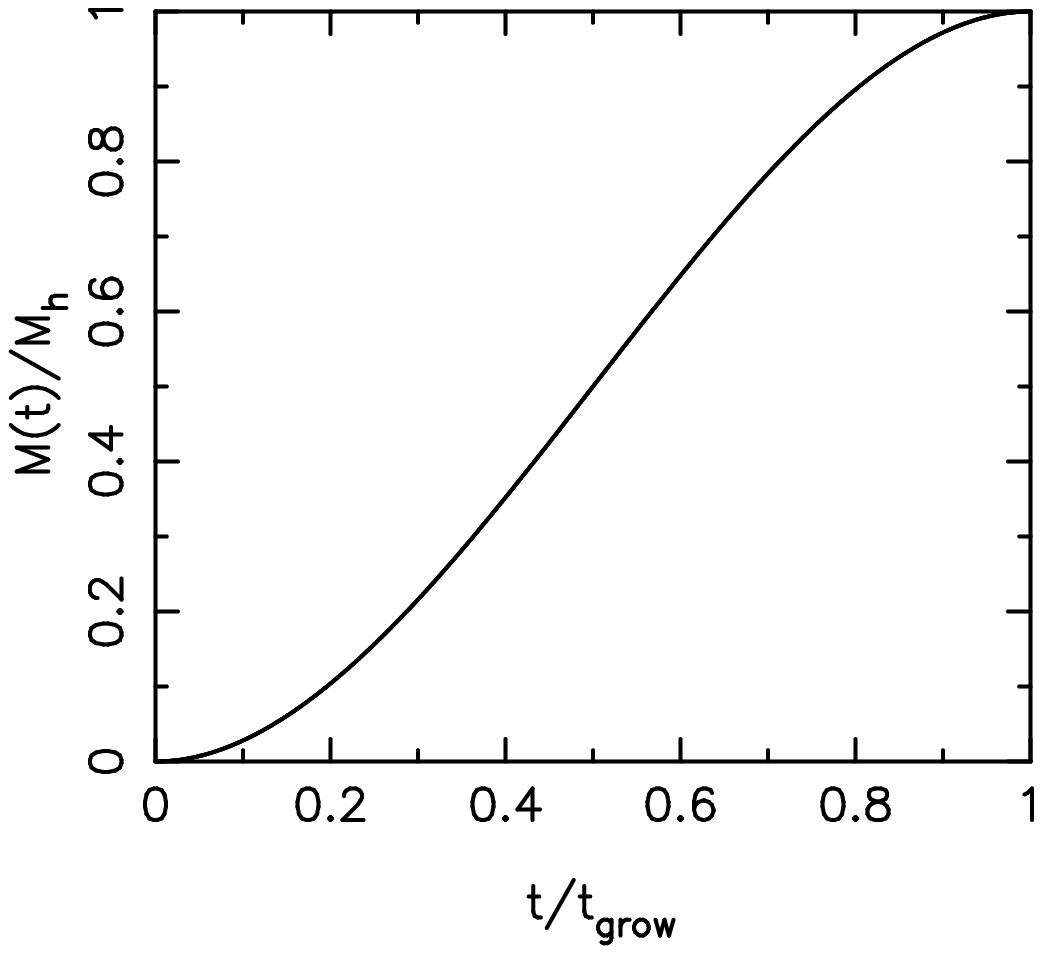]{\label{fig2}} 
The black hole mass as a function of time (Eq. 2).

\figcaption[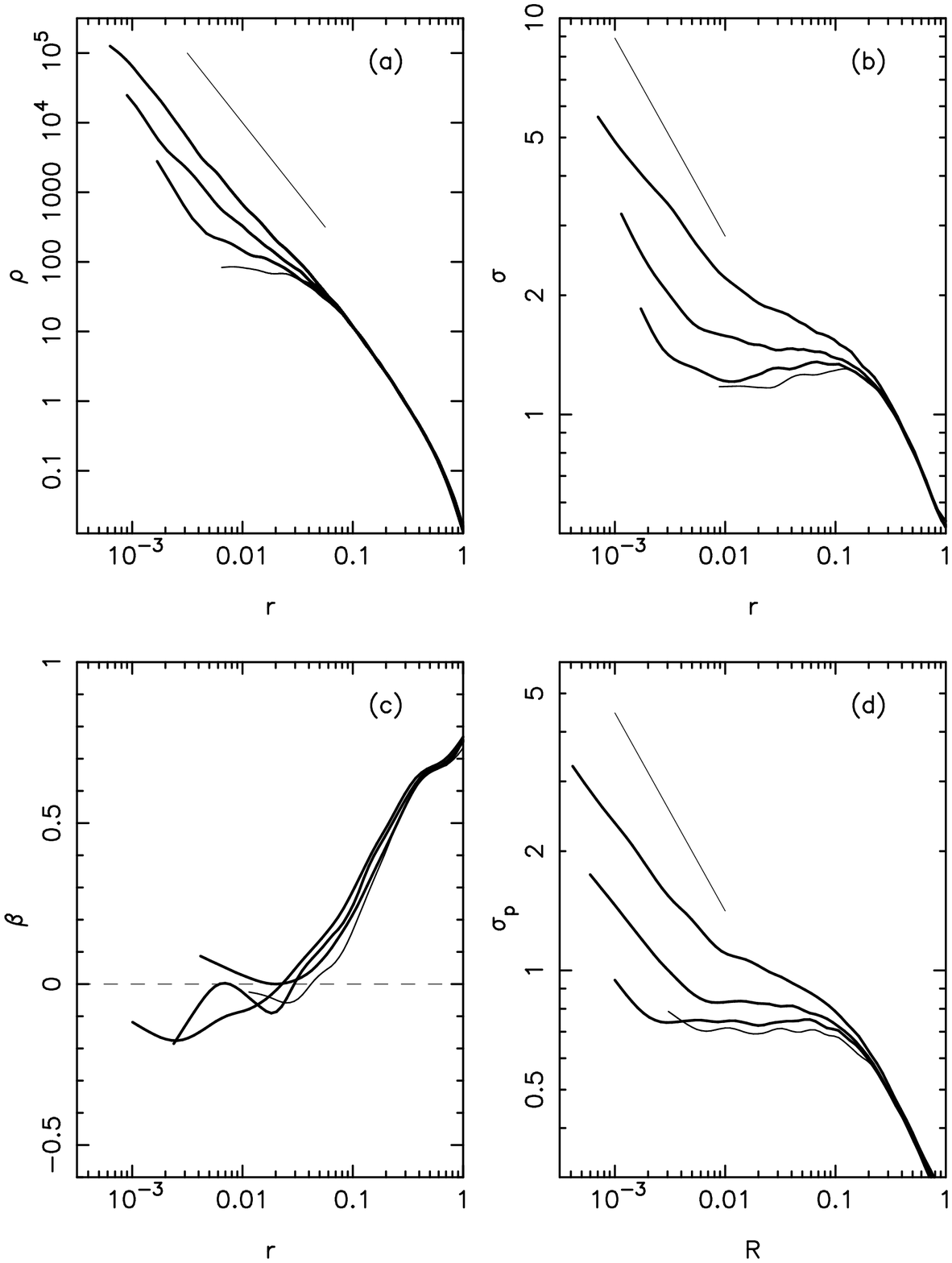]{\label{fig3}} 
(a)-(c) Spherically-averaged properties of the initial model 
(thin lines) and final models (heavy lines) in the integrations 
with $t_{grow}=15$, for $M_h=0.003, 0.1$ and $0.3$.
The leftward extent of the curves increases with $M_h$. 
(a) Space density; reference line has a logarithmic slope of
$-2$.
(b) One-dimensional velocity dispersion; reference line has a 
logarithmic slope of $-1/2$.
(c) Anisotropy parameter, $\beta=1-\sigma_t^2/\sigma_r^2$.
(d) Projected velocity dispersion, as seen from the direction 
of the intermediate axis; reference line has a logarithmic slope 
of $-1/2$.

\figcaption[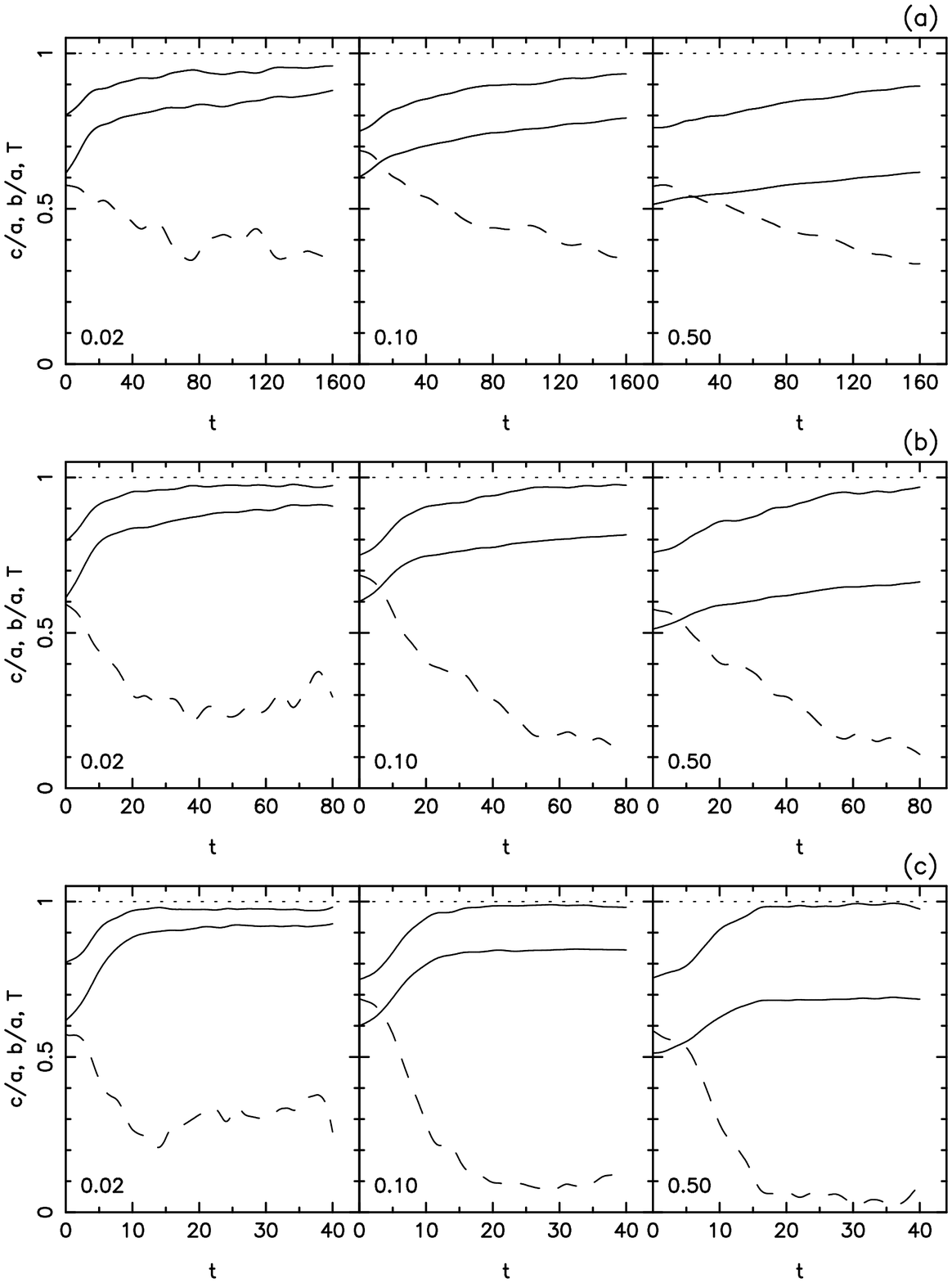]{\label{fig4}} 
Evolution of the axis ratios of the $N$-body models for 
$t_{grow}=15$. (a) 
$M_h=0.003$; (b) $M_h=0.01$; (c) $M_h=0.03$.
Solid lines: $b/a$ and $c/a$; dashed lines: triaxiality index, 
$T=(a^2-b^2)/(a^2-c^2)$.
The numbers in the lower left corner of each frame are the 
fraction of particles, ranked by binding energy, that were used 
in computing the axis ratios.

\figcaption[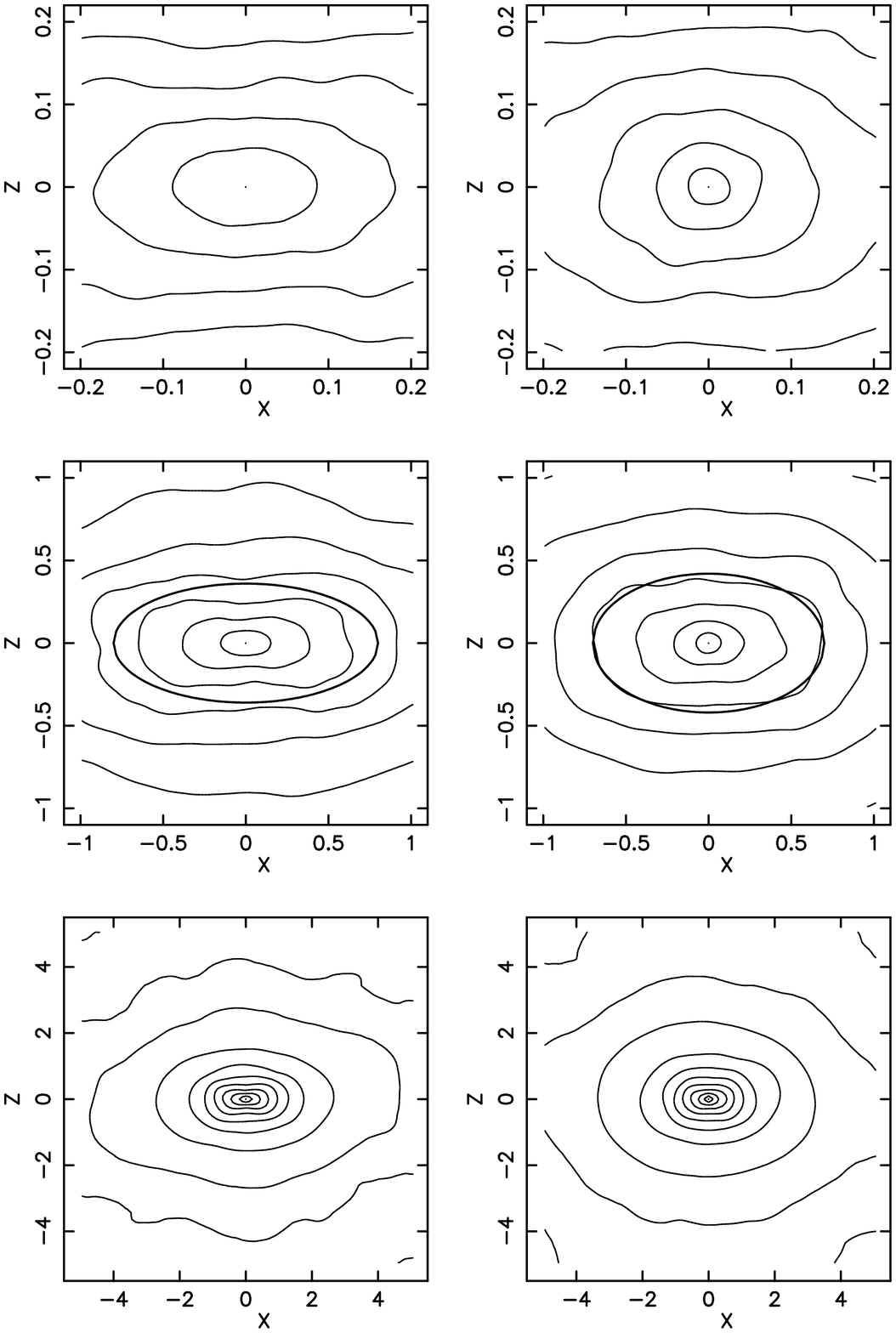]{\label{fig5}} 
Contours of the surface mass density as seen from the direction 
of the intermediate axis.
Left column: initial model; right column: final time step ($t=80$) 
of the model with $M_h=0.01$.
Contours are separated by 0.3 dex in the top row, and by 0.5 dex in 
the bottom two rows.
The heavy ellipse on the left has an axis ratio of 0.45; the ellipse
on the right has an axis ratio of 0.6.

\figcaption[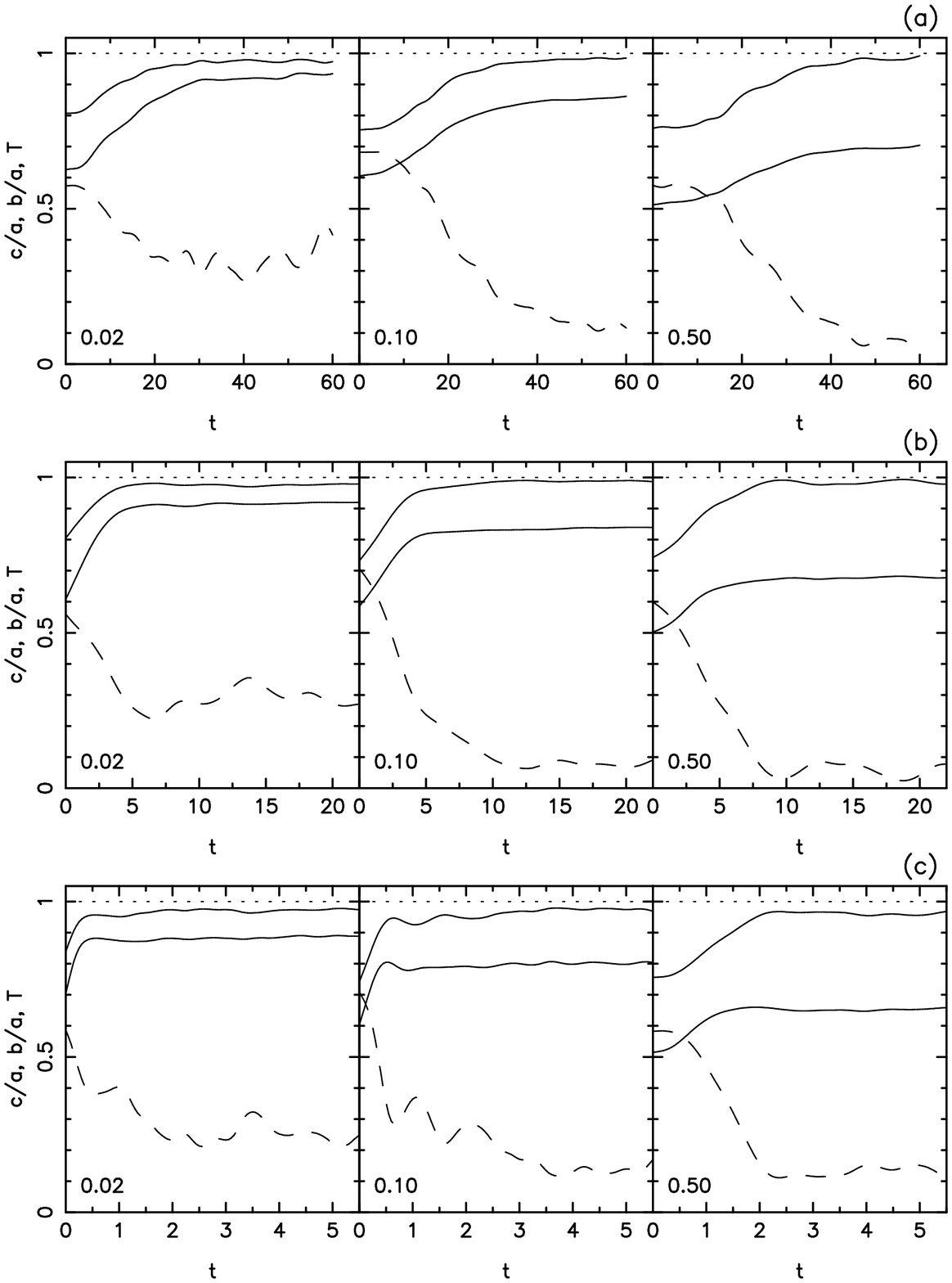]{\label{fig6}}
Evolution of the axis ratios of the $N$-body models with $M_h=0.03$. 
(a) $t_{grow}=45$; (b) $t_{grow}=5$; (c) $t_{grow}=0.1$.
Solid lines: $b/a$ and $c/a$; dashed lines: triaxiality index, 
$T=(a^2-b^2)/(a^2-c^2)$.
The numbers in the lower left corner of each frame are the 
fraction of particles, ranked by binding energy, that were used 
in computing the axis ratios.

\figcaption[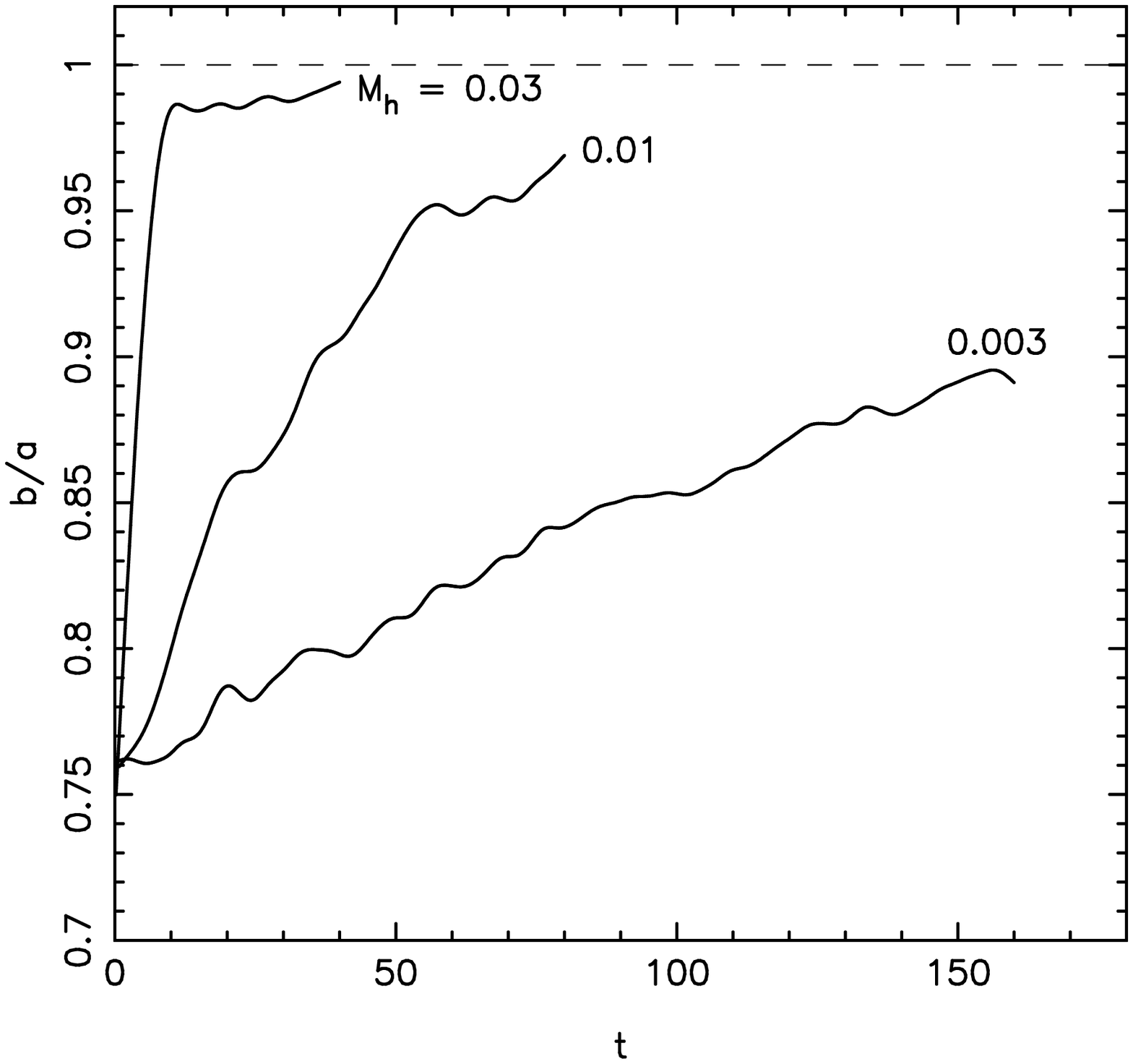]{\label{fig7}}
Evolution of the intermediate-to-long axis ratio $b/a$ defined by
the most-bound 50\% of the stars, for $M_h=0.003,0.01,0.03$.
The growth time of the black hole was $t_{grow}=15$ for the two 
smaller values of $M_h$, and $t_{grow}=5$ for $M_h=0.03$. 

\figcaption[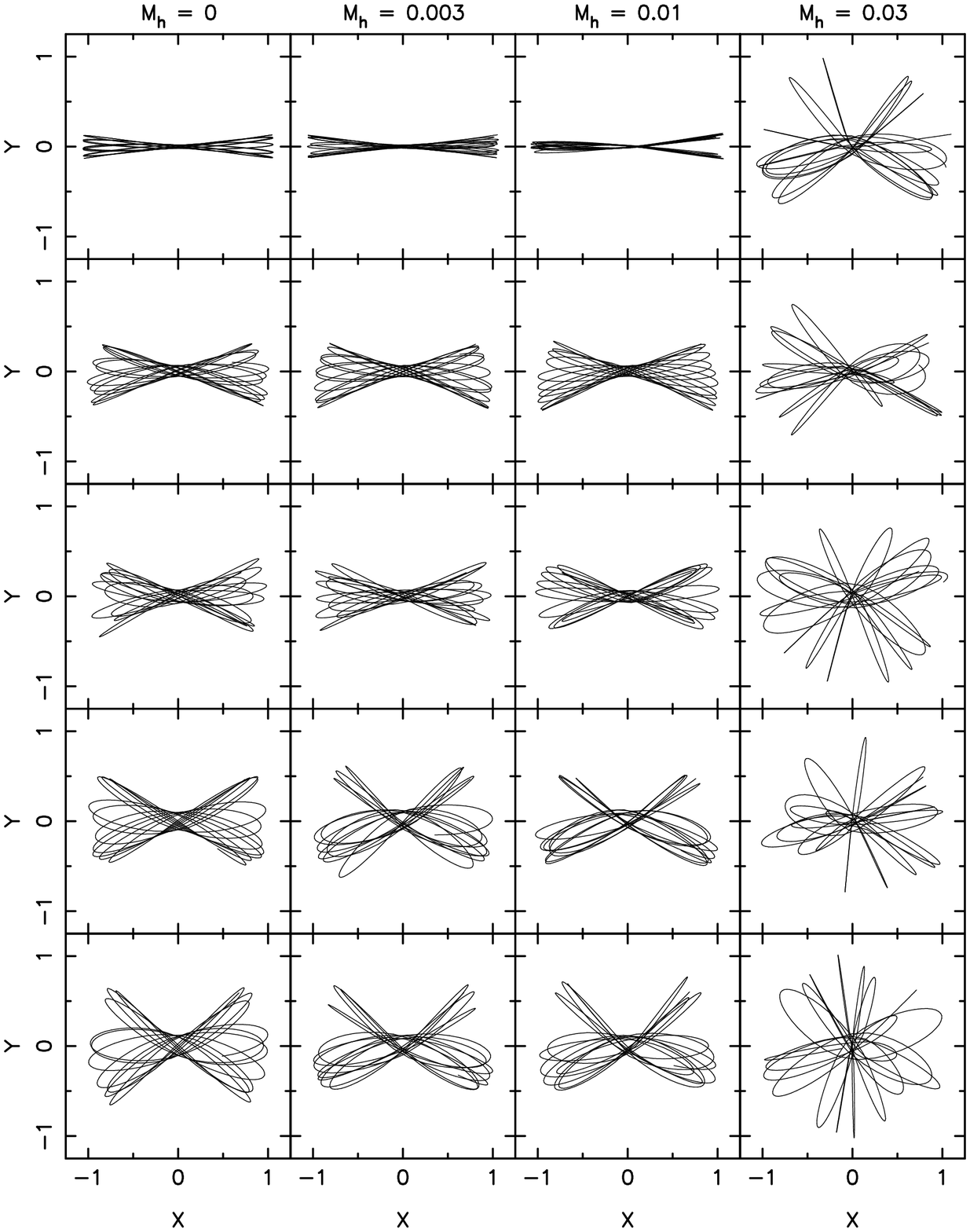]{\label{fig8}} 
Test-particle integrations in a triaxial model with a smooth core 
(Eq. 4) and a central point mass containing a fraction $M_h$ of 
the total mass of the model.
The $z$-axis is the short axis of the figure.
Each of the five orbits is defined by its starting point on the 
equipotential surface, which remained fixed in angular position 
as $M_h$ was increased.
Orbits were integrated for approximately 10 full oscillations.

\figcaption[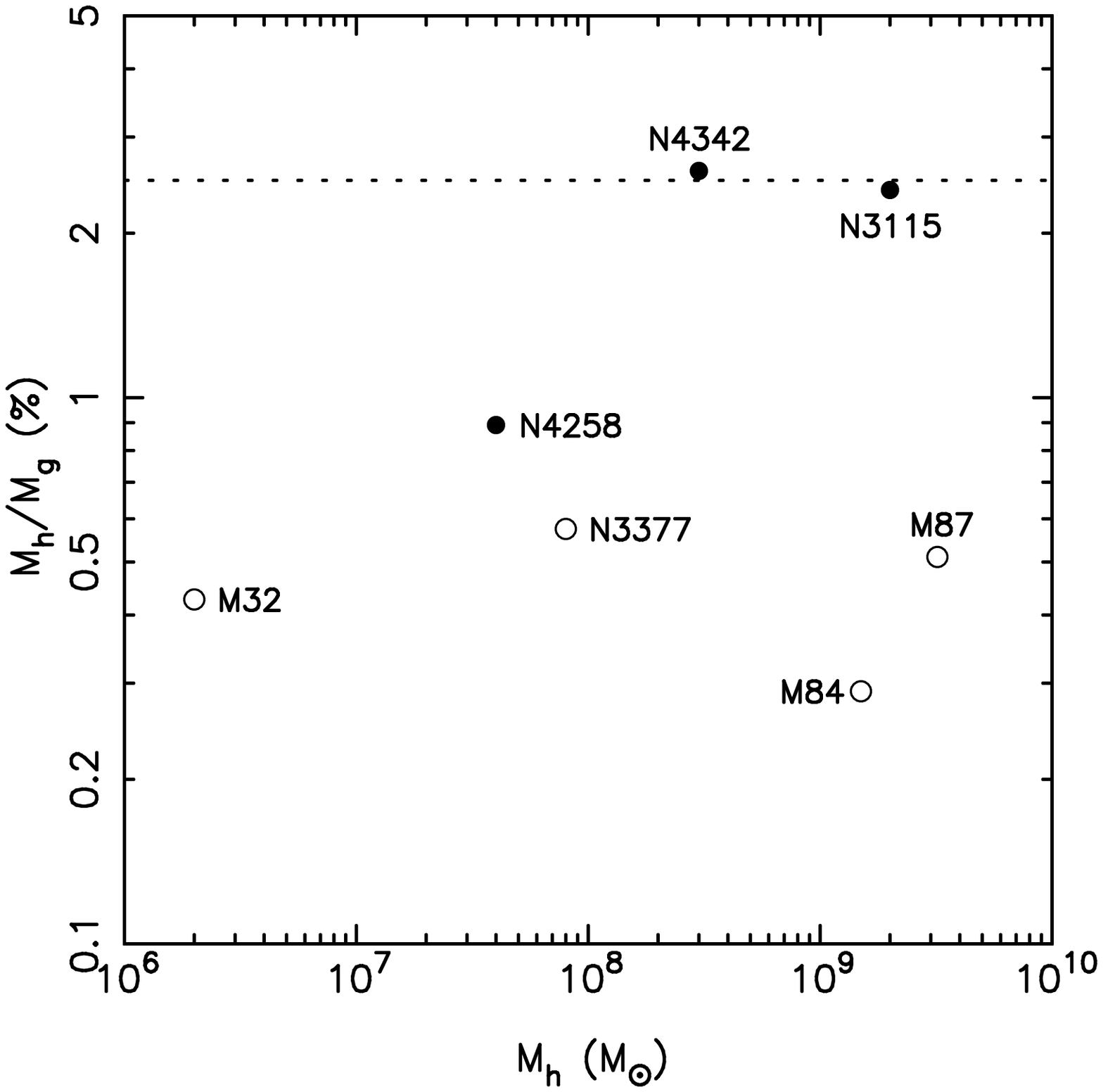]{\label{fig9}} 
Black hole masses $M_h$, in units of the solar mass, and mass fractions 
$M_h/M_g$ for seven galaxies with securely-detected black holes and
large black hole mass fractions 
(\cite{kor92}; \cite{mim95}; \cite{kor96};  
\cite{bow97}; \cite{mac97}; \cite{vaz97}; \cite{vdb97}).
$M_g$ is defined as the mass of the stellar bulge in the disk
galaxies (filled circles: NGC 4342, NGC 3115, NGC 4258), 
and as the total stellar mass in the elliptical galaxies 
(open circles: M32, M84, M87, NGC 3377).
The dashed line is the proposed upper limit to $M_h/M_g$.

\clearpage

\begin{table}
\caption{Properties of the $N$-Body Models}
\medskip
\begin{tabular}{cccccc} \hline
	    & & Initial      &	         & $M_h$  & 	   \\ 
Parameter   & & Model & $0.003$ & $0.01$ & $0.03$  \\ \hline

${M_{bnd}}^1$ & &   0.89 	    &	      &	       &	  \\
${T_{max}}^2$ & &		    &160      &80      &40  	  \\
${\epsilon}^3$& &--- &$4.7\times10^{-5}$&$1.6\times10^{-4}$ & $4.7\times10^{-4}$\\
${r_h}^4$     & &--- &$2.1\times10^{-3}$&$6.9\times10^{-3}$ & $2.1\times10^{-2}$\\
${r_{2}}^5$   & &0.046&0.041	        &0.034	        &0.023	    \\ 
$r_{10}$      & &0.11&0.10	        &0.094	        &0.080	    \\
$r_{50}$      & &0.48&0.47	        &0.45	        &0.43	    \\ 
${T_{2}}^6$   & &0.43&0.37	        &0.27	        &0.15       \\
$T_{10}$      & &0.70&0.64	        &0.58	        &0.45       \\
$T_{50}$      & &2.92&2.84	        &2.71	        &2.50       \\ \hline
\end{tabular}
\medskip

$^1$ Total mass of stars remaining bound to the initial model

$^2$ Integration time

$^3$ Softening radius of the black hole (eq. 3)

$^4$ $r_h=GM_h/\sigma_o^2$, where $\sigma_0$ is the one-dimensional
central velocity dispersion of the initial model

$^5$ $r_N = $ spherical radius containing $N\%$ of the total 
mass, measured at $t=40$ in the models with nonzero $M_h$

$^6$ $T_N = $ orbital period at radius $r_N$

\end{table}

\setcounter{figure}{0}

\begin{figure}
\plotone{figure1.ps}
\caption{ }
\end{figure}

\begin{figure}
\plotone{figure2.ps}
\caption{ }
\end{figure}

\begin{figure}
\plotone{figure3.ps}
\caption{ }
\end{figure}

\begin{figure}
\plotone{figure4.ps}
\caption{ }
\end{figure}

\begin{figure}
\plotone{figure5.ps}
\caption{ }
\end{figure}

\begin{figure}
\plotone{figure6.ps}
\caption{ }
\end{figure}

\begin{figure}
\plotone{figure7.ps}
\caption{ }
\end{figure}

\begin{figure}
\plotone{figure8.ps}
\caption{ }
\end{figure}

\begin{figure}
\plotone{figure9.ps}
\caption{ }
\end{figure}

\end{document}